\documentclass{an}
\usepackage{graphicx}
\usepackage{times}
\usepackage{fancyhdr}
\usepackage{amsmath}
\usepackage{amssymb}
\usepackage{psfig}
\newcommand{\lala}{$\lambda\lambda$}
\newcommand{\kms}{km s$^{-1}$}

\begin{document}

\title{Outflows and shocks in compact radio sources}

\author{J. Holt\inst{1}\fnmsep\thanks{Email: j.holt@sheffield.ac.uk }
\and  C. N. Tadhunter\inst{1}
\and  R. Morganti\inst{2}}
\institute{Department of Physics and Astronomy, University of Sheffield, Hicks Building, 
Hounsfield Road, Sheffield. S3 7RH. UK
\and
ASTRON, PO Box 2, 7990 AA Dwingeloo, The
Netherlands.}

\date{Received; accepted; published online}

\abstract{
We report some key results from the optical emission line
  study of a complete sample of compact radio sources. We find strong
  evidence for jet-driven outflows in the circum-nuclear emission line
  gas namely: 1) highly broadened and blueshifted emission line
  components (up to 2000 \kms), 2) shock ionised gas (broader, shifted
  components), 3) 
  consistency in the scales of the emission line gas and the radio
  source and 4) trends between the maximum outflow velocity and radio
  source size (and orientation).  Full details can be
  found in Holt  (2005).  
\keywords{galaxies: active; galaxies: ISM; ISM: jets and outflows; ISM: kinematics and dynamics. }}

\correspondence{j.holt@sheffield.ac.uk}

\maketitle

\section{Introduction}
\begin{center}
\begin{figure*}
\begin{center}
\begin{tabular}{cc}
\psfig{file=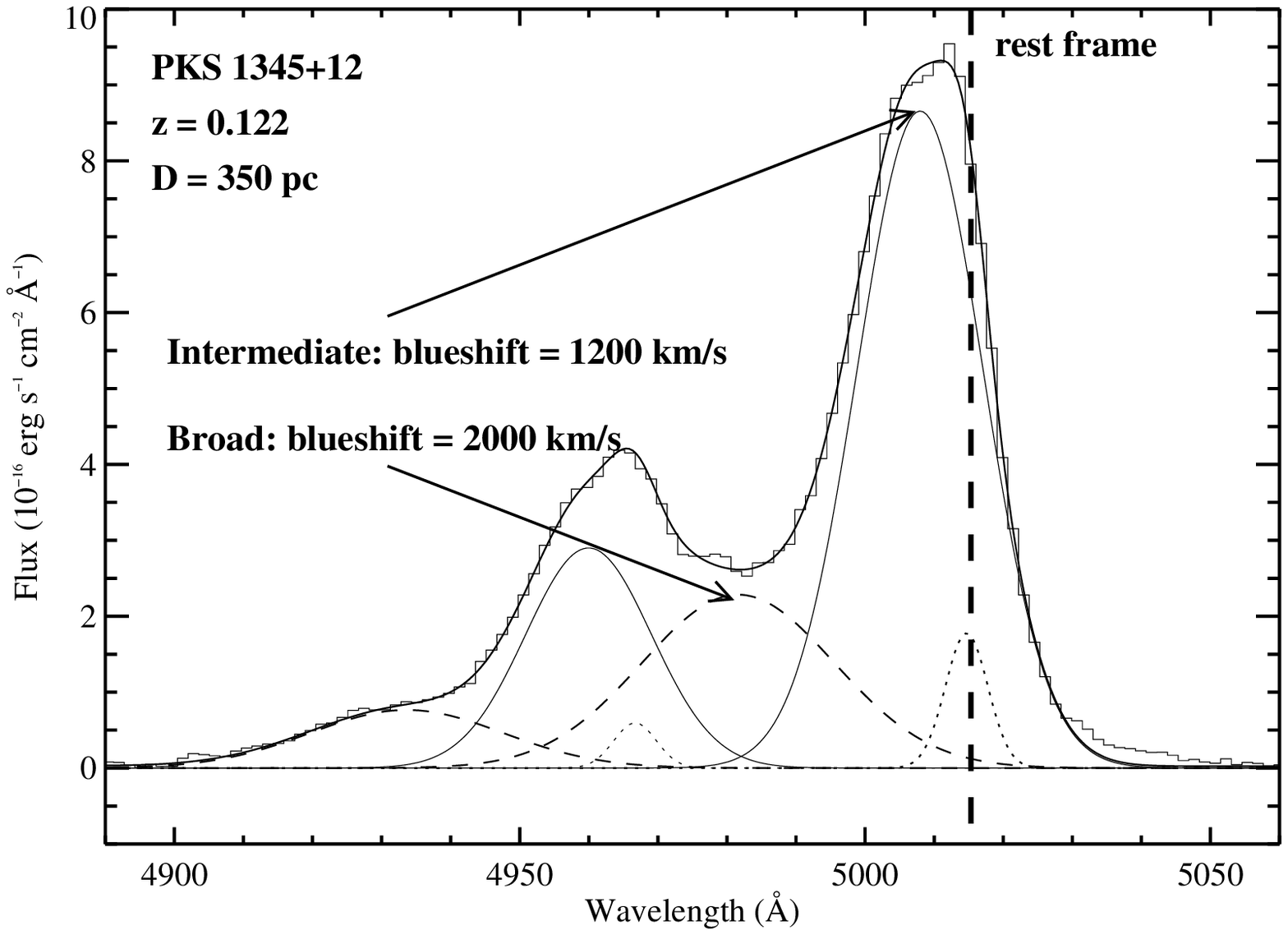,width=7.8cm,angle=0.} &
\psfig{file=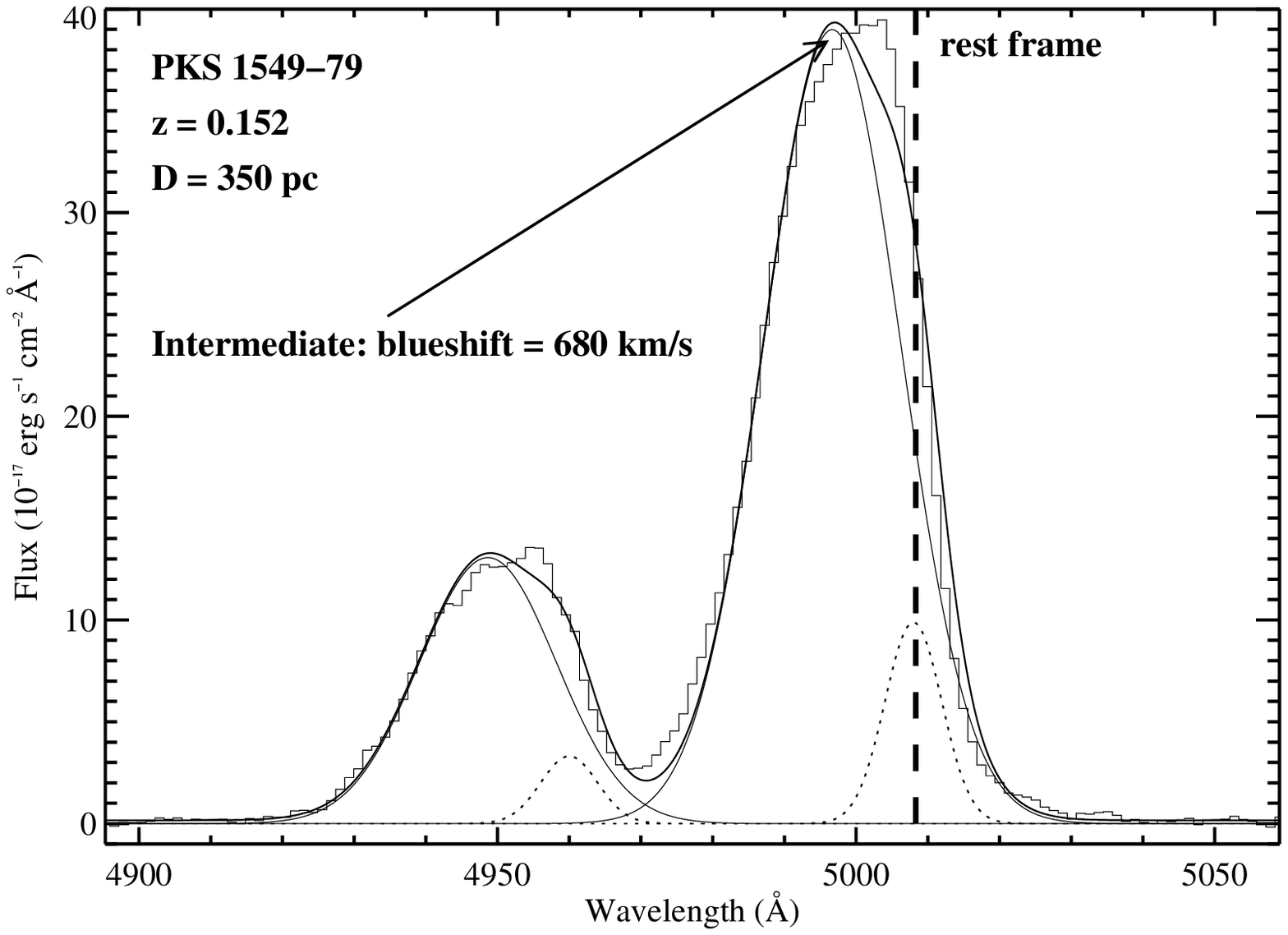,width=7.8cm,angle=0.} \\
\psfig{file=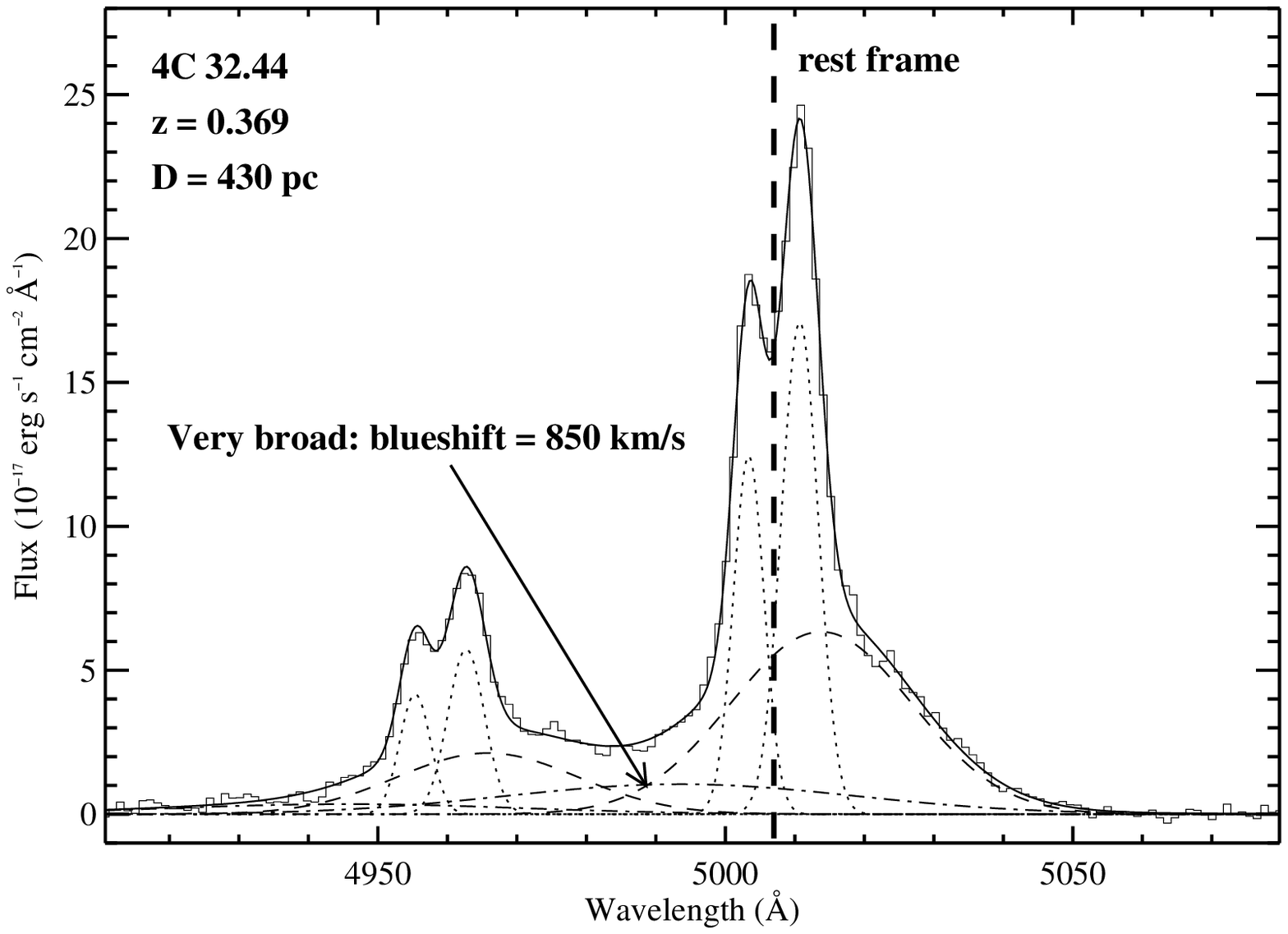,width=7.8cm,angle=0.} &
\psfig{file=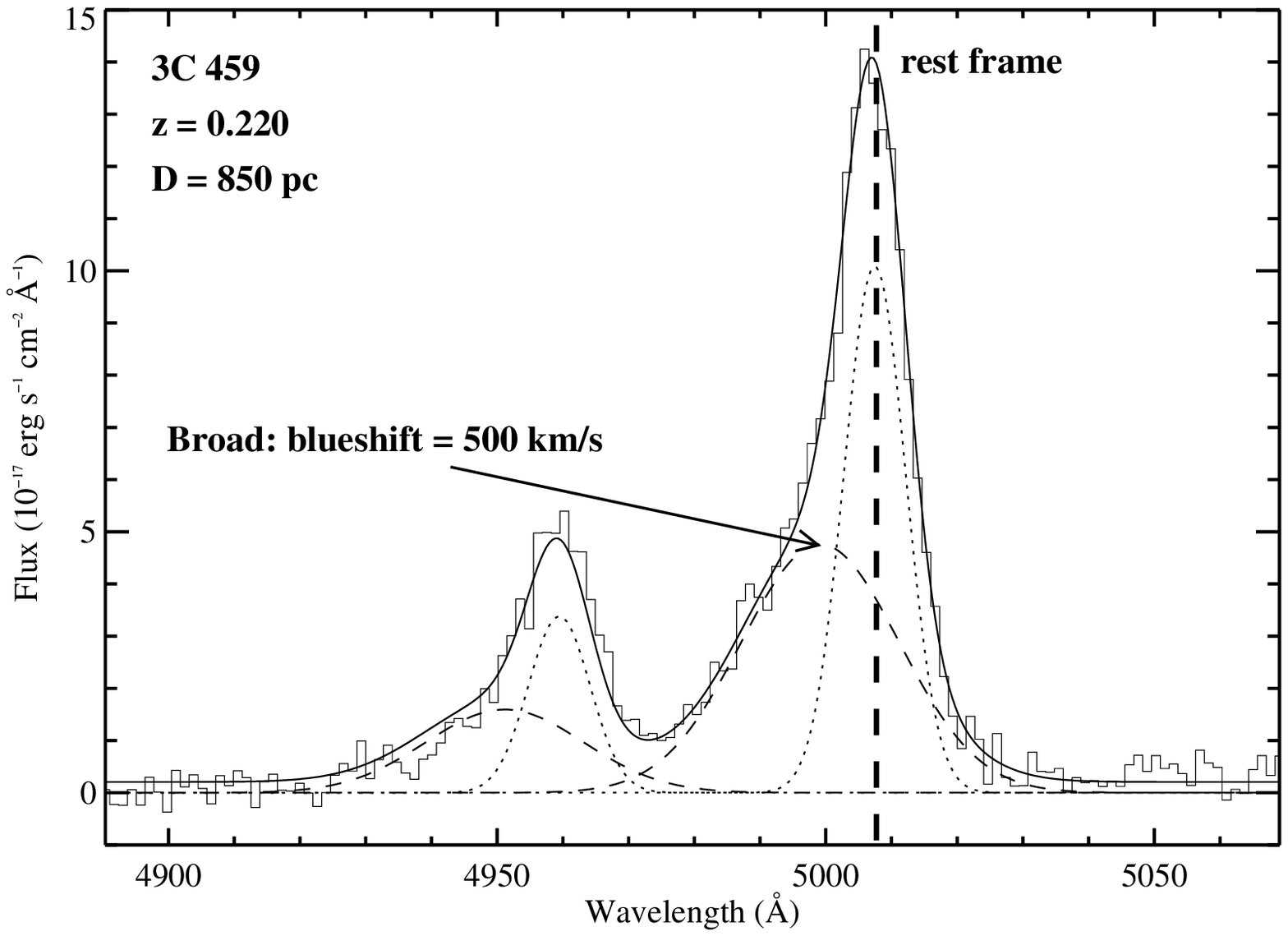,width=7.8cm,angle=0.} \\
\end{tabular}
\caption{Models for the {[O III]}\lala4959,5007 emission line doublet
  in four radio sources. For all, the faint line represents the data
  and the bold line traces the overall model. Different velocity
  components are as follows: narrow (FWHM $<$ 600 \kms: dotted),
  intermediate (600 $<$ FWHM $<$ 1200 \kms: solid), 
  broad (1200 $<$ FWHM $<$ 2000 \kms: dashed) and very broad (FWHM $>$
  2000 \kms: dot-dashed). The vertical line indicates the 
  the rest frame velocity. D is the radio source size. Note the
  unusually broad emission lines and the large asymmetries with strong
  blue wings.
}
\label{o3profiles}
\end{center}
\end{figure*}
\end{center}
Gigahertz-Peaked Spectrum Radio Sources (GPS: D $<$ 1 kpc) 
and the larger Compact Steep Spectrum Radio Sources (CSS: 
D $<$ 15 kpc) account for a significant fraction of the 
radio source population ($\sim$40\%) although their 
nature is not fully understood (see e.g. O'Dea 1998 and 
references therein). Currently, we believe they are {\em 
young} radio sources (Fanti et al. 1995) supported by 
estimates of dynamical ages: t$_{dyn}$ $\sim$ 
10$^{2}$-10$^{3}$ years (Owsianik et al. 1998); and radio 
spectral ages: t$_{sp}$ $<$ 10$^4$ years (Murgia et al. 
1999). This is in preference to the {\em frustration} 
scenario where the ISM is so dense, the radio jets cannot 
escape and the radio source remains confined and 
frustrated for its entire lifetime (van Breugel 1984). 

If compact radio sources are young, we will observe them 
relatively recently after the event(s) which triggered 
the activity (e.g. a merger; Heckamn et al. 1986). Hence, 
the circumnuclear regions will still retain large 
amounts of gas and dust deposited by the activity 
triggering event. 

During the early lifetime of the radio source, the young 
small scale radio jets will be on the same scale as this 
circumnuclear ISM and so will readily interact with it. 
Hence, one would expect to observe signatures of this 
interaction namely outflows in the emission line gas and 
evidence for jet-cloud interactions in the emission 
line ratios.

Indeed, Tadunter et al. (2001) reported evidence for fast 
outflows in the emission line gas in the compact flat 
spectrum radio source PKS 1549-79. From their low resolution optical
spectra, the high ionisation emission lines (e.g. {[O III]}) were both
broader (FWHM $\sim$ 1350 \kms~compared to $\sim$ 650 \kms) and {\it
  blueshifted} by $\sim$ 600 \kms~with respect to the low ionisation
lines (e.g. {[O II]}). Tadhunter et al. interpreted these unusual
  kinematics as the signature of the young small scale radio jets
  expanding out through a dense circumnuclear cocoon of gas and dust
  giving rise to outflows in the highly ionised emission line gas (see
  Figure 2 in Tadhunter et al. 2001). More recently, an extreme
  emission line   outflow (up to 2000 \kms) in the GPS source PKS
  1345+12 was reported by Holt, Tadhunter \& Morganti (2003).

Hence, we have obtained intermediate resolution 
(4-6\AA) optical spectra with good signal-to-noise over a 
large spectral range (with the WHT, NTT and VLT)
to search for such outflows in a 
statistically complete sample of 14 compact radio 
sources including 8 CSS, 3 GPS, 2 compact flat spectrum and 1 compact
core radio sources (see Holt 2005 for details). Here, we summarise some of the
main results   from this study. \\

\section{PKS 1345+12: the most extreme outflow}
As discussed in detail by Holt et al. (2003), PKS 1345+12 contains an
extreme emission line outflow with two outflowing components -- an
intermediate component (FWHM $\sim$ 1200 \kms) blueshifted by $\sim$
400 \kms~and a broad component (FWHM $\sim$ 2000 \kms) blueshifted by
$\sim$ 2000 \kms~with respect to the narrowest (FWHM $\sim$ 350 \kms)
component. The top-left panel in Figure \ref{o3profiles} shows the
highly complex {[O
    III]}\lala4959,5007 emission lines in the nuclear aperture of PKS
1345+12 and the components required to model the doublet.

By modelling the highly extended (up to $\sim$ 20 kpc)
emission line gas, the nuclear narrow component was shown to be
consistent with the rest frame of the galaxy. Hence, the broader
components trace blueshifted material flowing towards the
observer. Through reddening arguments, Holt et al. (2003) argue that
this material is on the side of the nucleus closest to the observer
and hence traces an outflow in the emission line gas. Indeed, this
result is supported by the observation of corresponding velocity
components in HI absorption by Morganti et al. (2003). 

\section{Emission line outflows}
In addition to the extreme emission line outflow observed in PKS
1345+12, fast emission line outflows are observed in 11 of the 14
compact radio sources in the sample. Three further examples of highly
complex   {[O III]} 
profiles are shown in Figure \ref{o3profiles}. 

\begin{figure}
\centerline{\psfig{file=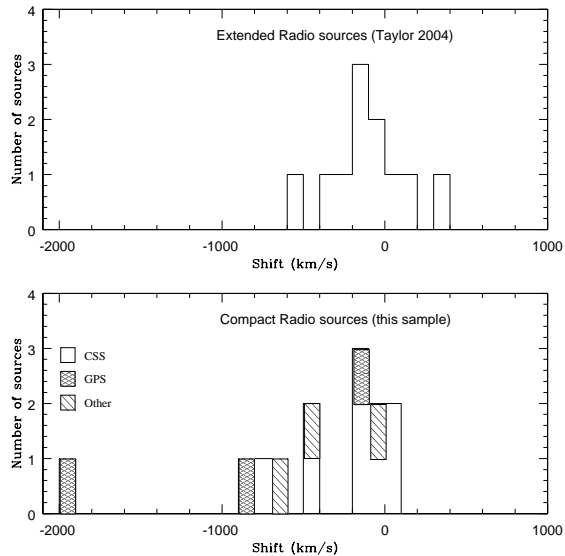,width=8cm,angle=0.}}
\caption{Histogram comparing the maximum outflow velocities in this
  sample of compact radio sources and a sample of extended radio sources.
}
\label{hist-ext-comp}
\end{figure}

To study the significance of the outflows statistically, we 
define the {\it maximum outflow velocity} to be the velocity shift between
the systemic velocity (often the narrowest component or, if there are
two narrow components, between the two narrow components) and the
broadest component taken from the emission line modelling (see Figure
\ref{o3profiles}). Figure \ref{hist-ext-comp} shows a pair of
histograms comparing the distribution of outflow velocities in this
sample of compact radio sources with a complete sample of extended radio
sources taken from Taylor (2004).

The distributions are clearly different with the compact radio sources
containing more extreme outflows than their extended
counterparts. Indeed, this trend is also evident within the sample of
compact radio sources -- the two highest 
outflow velocities are observed in some of the smallest (GPS) radio
sources. 
The distributions were tested using a Kolmogorov-Smirnoff
test and found to be different at the 99.9\% confidence level. Hence,
the size of the radio source is clearly important in determining the
outflow velocity of the emission line gas.

\begin{figure}
\centerline{\psfig{file=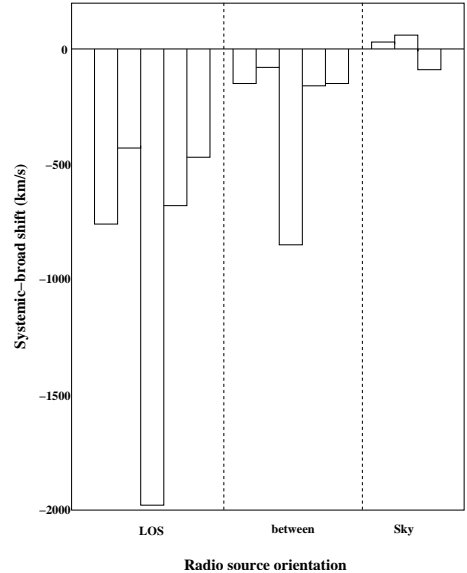,width=6cm,angle=0.}}
\caption{Histogram showing the effect of radio source `orientation' on
  the observed outflow velocity for the sample. }
\label{hist-orientation}
\end{figure}

A more tenuous result links the observed outflow velocity to the
orientation of the radio source. Higher outflow velocities are
generally 
observed in radio sources orientated close to the observer's
line of sight whilst radio sources close to the plane of the sky tend
to have much smaller outflow velocities.  
Note, however, for the majority of the sources it
was impossible to determine accurate orientations from the radio maps
available. However, as orientation is likely to be important to some
degree,  the sample was divided into three broad
categories (close to the line of sight, close to the
plane of the sky  and `in between')
using the radio map symmetry/asymmetry and whether a radio
core was detected or not. 
The results are plotted in Figure
\ref{hist-orientation}. 

Finally, two radio sources (PKS 1345+12 and PKS 1549-79) were imaged
using ACS on the HST (see Tadhunter et al. 2005, in prep.) revealing
the bright emission line regions to be on similar scales to the radio
source. Hence, these {\it kinematic} results are consistent with the
idea that the young, small scale radio sources expand through an
enshrouding cocoon giving rise to outflows in the emission line gas.

\section{Ionisation mechanisms - the evidence for shocks}

\begin{figure*}
\begin{center}
\begin{tabular}{c}
\psfig{file=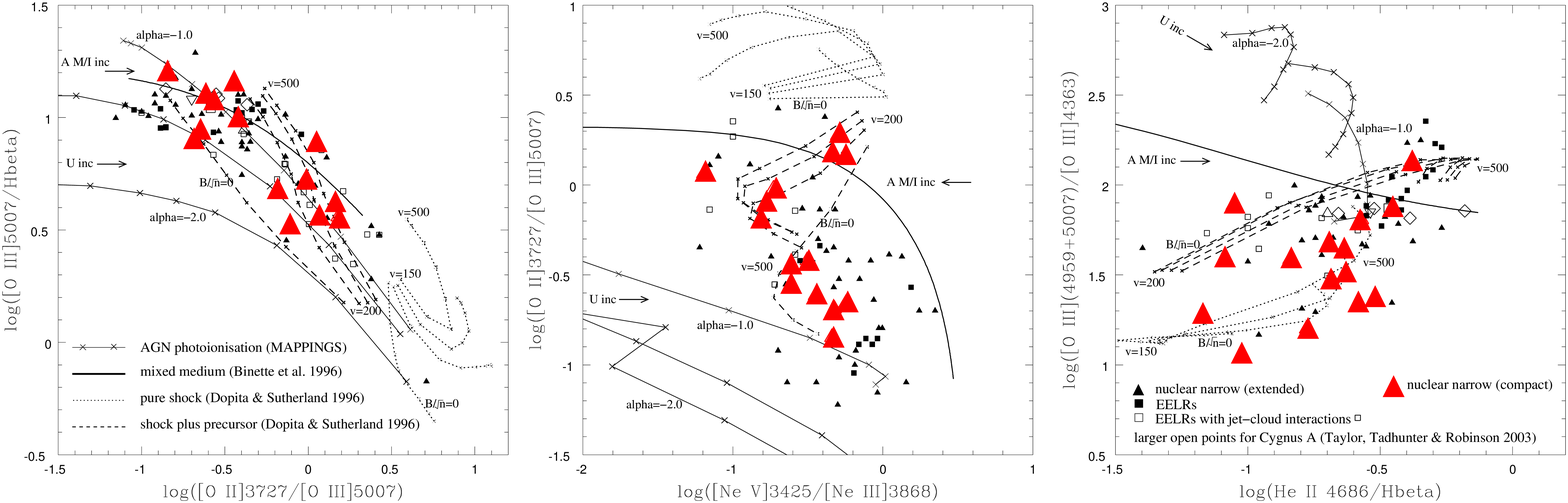,width=17cm,angle=0.}\\
\psfig{file=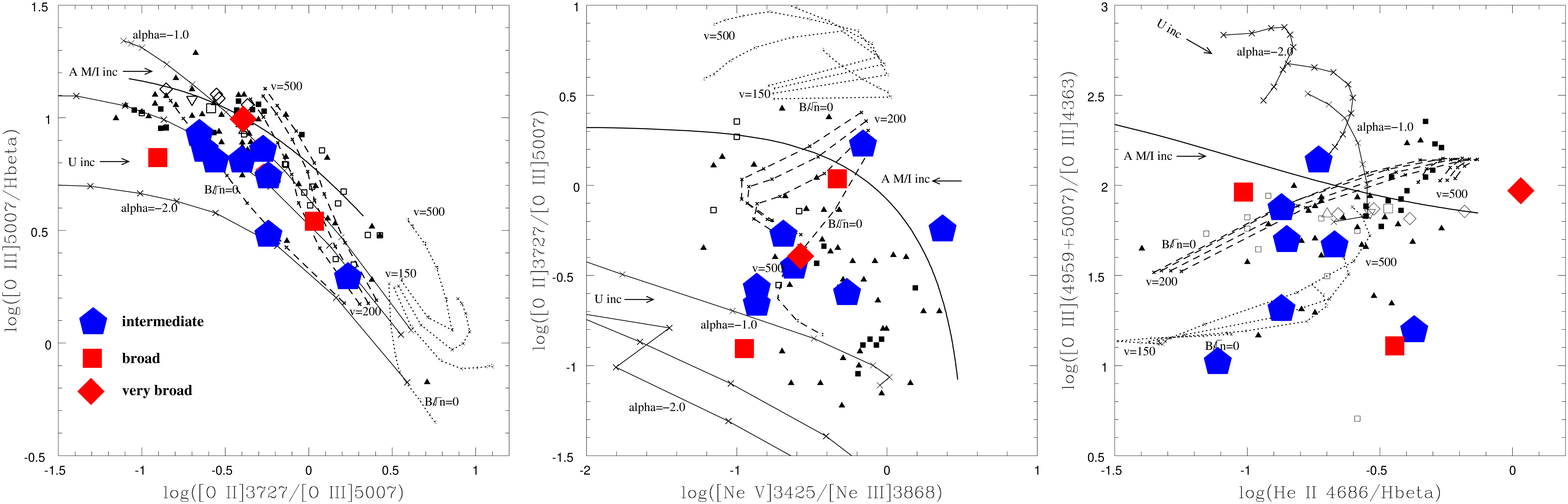,width=17cm,angle=0.}\\
\end{tabular}
\caption{A selection of diagnostic diagrams for the nuclear narrow
  components (top row) and the nuclear shifted components (bottom
  row). A full discussion is presented in Holt (2005). The data are
  compared to theoretical models including: \newline (1)
AGN-photoionisation calculated using {\sc mappings}. Each line traces
  a sequence in U, the ionisation parameter (2.5 $\times$ 10$^{-3}$
  $< U <$ 10$^{-1}$) for given values of $\alpha$ (-2.0, -1.5 and
  -1.0) where F$_{\nu}$ $\propto$ $\nu^{\alpha}$.\newline
(2) mixed medium photoionisation including both ionisation and matter
  bounded clouds: 10$^{-2}$ $<$ A$_{M/I}$ $<$ 10 (Binette et al. 1996).\newline
(3) shock ionisation (pure shocks and models including 50\% precursor)
  for shock velocities 150 $<$ v$_{\rm shock}$ $<$ 500 \kms~for
 a given magnetic parameter (B/$\sqrt{n}$ = 0,1,2,4 $\mu$G
  cm$^{3/2}$) taken from Dopita \& Sutherland (1996). 
The smaller points are for extended radio sources taken
  from the literature including nuclear regions, EELRs and EELRs with
  evidence for jet-cloud interactions (see Holt 2005 for details). The
  open points are for Cygnus A 
  taken from Taylor, Tadhunter \& Robinson (2003).
}
\label{ionisation}
\end{center}
\end{figure*}

Emission line ratios were used to search for further evidence of shock
ionisation, a common feature in the {\it extended} emission line
regions coincident with the radio source around
 some extended radio sources (e.g. Villar-Mart{\'{i}}n
et al. 1997,1998, Best et al. 2000, Sol{\'{o}}rzano-I{\~{n}}arrea et
al. 2001, Inskip et al. 2002). In contrast to previous studies, we
plot both a larger sample of compact radio sources and also use the
kinematic subcomponents rather than treating the lines as single
Gaussians. A selection of the diagnostic diagrams presented in Holt
(2005) for the nuclear narrow components and nuclear shifted
components is shown in Figure \ref{ionisation}. 

The nuclear narrow components are generally consistent with
photoionisation models and are split roughly equally between AGN
photoionisation and mixed medium models. However, the nuclear broader
components are generally consistent with fast shocks (v$_{\rm shock}$
$\geq$ 300 \kms), often with a strong precursor component. Further
evidence for shocks comes from the kinematical results (see above) and
the measurement of high temperatures (T$_{e}$ $\gtrsim$ 14,000 K) in
many sources.

Hence, the {\it nuclear shifted} components appear to show similar
characteristics to the {\it extended} emission line regions with
evidence for jet-cloud interactions in extended radio sources. 
At face value these results appear quite different. However, when the
scale of the radio source is taken into account, this further
strengthens the idea that compact radio sources and extended radio
sources are scaled versions of each other. In extended radio sources,
the radio source is on a large scale comparable to the scale of the
extended emission line regions, hence shocks are sometimes observed in
their EELRs. However, in compact radio sources, the radio jets are
small and on the scale of the {\it nuclear} regions and so, if shocks
are important, they will be observed in the nuclei of compact radio
sources.

\section{Conclusions}
Our results show convincing evidence for the scenario in which compact
radio sources are the young relatives of the extended radio
sources. As the small scale radio jets expand through the dense cocoon
of gas and dust deposited during the triggering event (most likely a
merger), they sweep it aside giving rise to outflows in the emission
line gas. This study has found several key signatures of this scenario
namely: 
\begin{itemize}
\item fast outflows (up to 2000 \kms) in the nuclear emission line gas (broad,
  blueshifted emission line components). 
\item shock ionised gas (broader shifted components).
\item consistency in the scales of the emission line gas and radio
  source.
\item trends between the maximum outflow velocity and radio source
  size and, more 
  tenuously, radio source orientation. \vspace*{11pt}\\
\end{itemize}

\acknowledgements
JH acknowledges a PPARC PhD studentship and PDRA. 
This work is based on observations taken using the William Herschel
Telescope, La Palma, Spain; the European Southern Observatory New
Technology Telescope, La Silla, Chile and ESO's Very Large Telescope,
Cerro Paranal, Chile.

\end{document}